# Advanced RAG Models with Graph Structures: Optimizing Complex Knowledge Reasoning and Text Generation

Yuxin Dong
Wake Forest University
Winston-Salem, USA

Shuo Wang
Purdue University
Indianapolis, USA

Hongye Zheng
The Chinese University of Hong Kong
Hong Kong, China

Jiajing Chen
New York University
New York, USA

Zhenhong Zhang
George Washington University
Washington, USA

Chihang Wang *
New York University
New York, USA

*Abstract*—This study aims to optimize the existing retrieval-augmented generation model (RAG) by introducing a graph structure to improve the performance of the model in dealing with complex knowledge reasoning tasks. The traditional RAG model has the problem of insufficient processing efficiency when facing complex graph structure information (such as knowledge graphs, hierarchical relationships, etc.), which affects the quality and consistency of the generated results. This study proposes a scheme to process graph structure data by combining graph neural network (GNN), so that the model can capture the complex relationship between entities, thereby improving the knowledge consistency and reasoning ability of the generated text. The experiment used the Natural Questions (NQ) dataset and compared it with multiple existing generation models. The results show that the graph-based RAG model proposed in this paper is superior to the traditional generation model in terms of quality, knowledge consistency, and reasoning ability, especially when dealing with tasks that require multi-dimensional reasoning. Through the combination of the enhancement of the retrieval module and the graph neural network, the model in this study can better handle complex knowledge background information and has broad potential value in multiple practical application scenarios.

I. INTRODUCTION

At present, generative models have shown great potential in many application fields [1]. However, traditional generative language models mainly rely on a large amount of pre-trained data for learning. Although these models can generate relatively coherent text, their generation ability is still limited when facing external knowledge or complex background information. To this end, the retrieval-augmented generation model (RAG) came into being. By combining the retrieval system with the generation model, it can dynamically access external knowledge during the generation process, thereby greatly improving the quality and accuracy of generation. Especially in tasks that require a rich knowledge background or complex reasoning, the RAG model shows stronger advantages than the single-generation model [2].

However, the existing RAG model still has certain limitations [3]. First, the retrieval module of the RAG model is usually text-based, which means that it is difficult to effectively process complex structured information (such as graph structure [4], hierarchical relationship [5], etc.). In practical applications, a lot of knowledge has inherent structured characteristics, especially in scenarios such as knowledge graphs [6], social networks, scientific literature, etc., where information is not just discrete text fragments, but has a high degree of association and contextual relationship [7-8]. Faced with this complex graph structure information, traditional RAG models are often difficult to fully utilize, thus affecting the generation effect. Secondly, the synergy between the retrieval module and the generation module still needs to be further optimized. Although the current RAG model can use external information to enhance the generation effect, it still has inaccurate or incoherent generation results when dealing with complex reasoning tasks, which limits its application in some high-demand scenarios [9].

In order to solve the above problems, the RAG model combined with graph structure has gradually become a hot topic of research. Graph structure data can express the complex relationship between different objects in the form of nodes and edges, so it is widely used in many fields, such as knowledge graphs [10], financial networks [11], and medical analysis.[12]. By introducing technologies such as graph neural networks (GNNs), RAG models can better process structured knowledge and extract valuable information from them for the generation process. For example, in the knowledge graph, each node represents an entity and the edge represents the relationship between entities. The graph neural network can effectively capture this association structure, so that the model can access richer contextual information when generating. This RAG model combined with graph structure can better cope with complex knowledge reasoning tasks and significantly improve the generation quality and rationality of the model.

In practical applications, the RAG model based on graph structure has a wide range of potential values. In the medical field, the graph RAG model can be used for medical record generation and diagnosis assistance. Medical records usually contain a variety of structured information such as the patient's historical condition, diagnosis, and treatment plan. The graph-based model can better handle the association between this information and generate reports or suggestions with diagnostic value [13]. In short, the optimization and application research of the graph-based retrieval enhancement generation model RAG has important practical significance and broad application prospects. By introducing graph structure information, the RAG model can demonstrate stronger expressiveness in processing complex knowledge reasoning and generation tasks. At the same time, with the continuous development of technologies such as graph neural networks, the graph RAG model will also usher in more optimization and application opportunities in the future, providing technical support for intelligent generation in more fields.

## II. RELATED WORK

This study enhances the Retrieval-Augmented Generation (RAG) model by incorporating graph structures through Graph Neural Networks (GNNs) to support complex reasoning and improve generation quality. Prior research has informed various aspects of this work, from foundational NLP improvements to methodologies for structuring and processing complex, interconnected data.

A central component of the RAG model is the transformer architecture, which has been instrumental in handling semantic complexity. Du et al. [14] discuss transformers' applications in managing intricate semantic information in natural language processing (NLP), highlighting the mechanisms that allow models to capture nuanced relationships within text. Such advancements are directly relevant to this study's aim to process complex graph structures within RAG models, enhancing retrieval capabilities and accuracy in text generation.

Several studies have contributed to understanding structured knowledge processing through GNNs and other embedding techniques. Wei et al. [15] propose a self-supervised GNN model that improves feature extraction across heterogeneous information networks, a methodology closely aligned with this paper's approach. Their techniques provide a foundation for integrating GNNs in RAG models, improving the model's ability to understand complex associations within structured knowledge, such as in knowledge graphs. Similarly, Yang et al. [16] demonstrated the potential of dynamic hypergraphs in managing sequences and associations within data, supporting the notion that graph-enhanced models can effectively capture interconnected information within knowledge-rich domains.

The model's retrieval module benefits from embedding strategies and attention mechanisms for feature enhancement, as highlighted by Liu et al. [17]. Their work on using separation embedding and self-attention strengthens understanding of how embeddings can improve interpretability and contextual sensitivity, both crucial for this paper's RAG model in ensuring knowledge consistency and coherence. Cang et al. [18] also explore ensemble methods with transformer-based models to handle specialized datasets, providing insight into techniques for augmenting model adaptability when working with structured, domain-specific information.

Another relevant contribution involves approaches for model optimization and improved robustness in handling structured data. Jiang et al. [19] developed a weighted adversarial network to enhance cross-domain consistency, a concept that aligns with the present study's objective to improve the accuracy and reliability of generated responses when handling multifaceted knowledge. Additionally, Xu et al. [20] applied deep learning architectures to predict sequential and temporally linked information, exemplifying effective strategies for embedding temporal relationships that mirror the graph structure's role in maintaining logical flow within generated text.

Moreover, the integration of interpretable data transformations, such as Yan et al.'s [21] transformation of multidimensional time series into interpretable sequences, underscores the value of structured data processing. These methods inform this study's approach to embedding graph structures in the RAG model to improve interpretability and coherence across generated content, particularly for complex reasoning tasks.

## III. METHOD

In this study, we proposed a graph-based retrieval-enhanced generation model (RAG) optimization scheme, and combined it with a graph neural network (GNN) to process graph structure information, thereby improving the model's generation ability in complex scenarios. To achieve this goal, our model framework mainly consists of three parts: graph structure processing module, retrieval module, and generation module as shown in Figure 1.

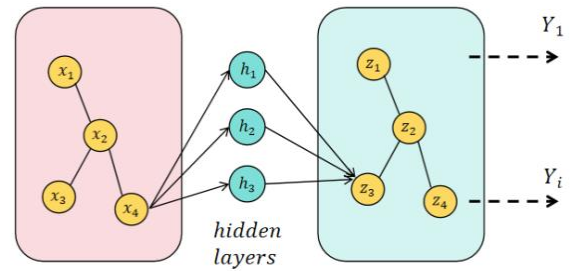

Figure 1 Graph network architecture

First, in the graph structure processing module, a graph neural network is used to encode the graph structure data. Assume that a graph structure can be represented as $G = (V, E)$, where V is a set of nodes and E is a set of edges. Each node $v \in V$ represents an entity or object, and edge

$e = (v_i, v_j)$ represents the relationship between nodes $v_i$ and $v_j$. For each node, we first represent its initial feature as $h_i^0$ and then update the representation of each node through the propagation mechanism of the graph neural network. The basic propagation mechanism of graph neural networks can be defined as:

$$h_i^{(k+1)} = \sigma(W^{(k)} \cdot AGG(\{h_j^{(k)} : j \in N(i)\}) + b^{(k)})$$

Among them, $N(i)$ represents the set of neighbor nodes of node $v_i$, AGG is the aggregation function, common aggregation methods include average, sum or maximum, $\sigma(\cdot)$ is the nonlinear activation function, $W^{(k)}$ and $b^{(k)}$ are the weight matrix and bias term of the kth layer respectively. After multiple layers of propagation and updating, we can get the final representation $h_i^{(L)}$ of each node, where L is the number of layers of the graph neural network.

Next, in the retrieval module, we perform retrieval in combination with graph structure information. In order to improve retrieval efficiency, we adopt a retrieval method based on graph embedding. Suppose we have a knowledge base K, which contains multiple knowledge fragments, each of which can be represented by a graph structure as $G_k$. We first perform graph encoding on each $G_k$ to obtain the embedding vector $z_k$ of each knowledge fragment. Then, for the input query, we also generate the graph embedding vector $z_q$ of the query through the graph neural network, and calculate the similarity between the query and each knowledge fragment in the knowledge base through the similarity between the vectors (cosine similarity):

$$sim(z_q, z_k) = \frac{z_q \cdot z_k}{\|z_q\| \|z_k\|}$$

The knowledge fragments with the highest similarity will be retrieved as the input of the generation module. This graph-based retrieval method can more effectively capture complex entity relationships and improve the accuracy of retrieval.

Finally, in the generation module, we use a generation model based on Chatgpt4.0 to generate text. Assuming that the retrieval module returns relevant knowledge fragments, each knowledge fragment is represented as a vector, we embed these retrieved knowledge fragments and input the query into the generation model together. During the generation process, the generation model will dynamically adjust the generation strategy according to the relevant information of these knowledge fragments to improve the relevance and accuracy of the generation results.

Specifically, the goal of the generation process is to maximize the conditional probability $P(y | q, z_1, z_2, ..., z_N)$ given a query $q$ and retrieved knowledge fragment $z_1, z_2, ..., z_N$, where $y$ represents the generated target text. This conditional probability can be represented by the output distribution of the generation model:

$$P(y_t | y_{<t}, q, z_1, z_2, ..., z_N) = soft\max(W_o \cdot h_t + b_o)$$

Among them, $y_t$ represents the t-th word generated, $y_{<t}$ represents the first t-1 words generated, $h_t$ is the hidden state of the generative model at time step t, and $W_o$ and $b_o$ are the parameters of the output layer. By continuously generating the next word until the generation end mark is reached, the model can complete the generation of the entire text.

In summary, this method introduces graph neural networks to process graph structured data and combines it with a retrieval-enhanced generation model to effectively utilize external knowledge in the generation process, thereby achieving better generation effects for complex knowledge scenarios.

## IV. EXPERIMENT

### A. Introduction to the dataset and the LLM used

In this experiment, we used a generative model based on ChatGPT-4 as the core part of the experiment. ChatGPT-4 is a large-scale pre-trained language model developed by OpenAI. It is based on the Transformer architecture and can generate natural and coherent text [22]. ChatGPT-4 has strong language understanding and generation capabilities and can handle a wide range of text-generation tasks. However, in order to further improve the performance in knowledge-intensive and reasoning tasks, we combined the design of the RAG (Retrieval Augmented Generation) model to dynamically obtain relevant information from external knowledge bases through the retrieval module to generate more knowledge-accurate and relevant answers. This architecture can effectively make up for the problem that the generative model relies solely on training data, enabling it to handle knowledge reasoning and complex tasks in more fields.

In our experiments, we used the Natural Questions (NQ) dataset, which is widely used for retrieval and generation tasks. NQ is an open-domain question-answering dataset that contains real question queries from users, accompanied by relevant document snippets automatically retrieved from Internet resources such as Wikipedia. Each data sample consists of a query, a retrieved document, and an answer, covering a wide range of content in multiple fields such as history, science, and culture. The diversity of the NQ dataset makes it an ideal choice for evaluating the performance and reasoning ability of generative models on complex domain problems. This experiment uses this dataset to test the retrieval and generation capabilities of our model to evaluate the accuracy and knowledge completeness of its generated text.

This experiment is run in a high-performance computing environment. The experimental hardware configuration includes NVIDIA A100 GPU, 128GB memory and 64-core CPU, which can support efficient training and reasoning of large-scale models. The experimental software is mainly based on the PyTorch framework for model training and evaluation, and combines Hugging Face's Transformers library to implement the retrieval and generation functions of the RAG model.

*B. Experimental Results*

We will select five different generative models to conduct comparative experiments with the RAG model based on graph structure optimization proposed in this paper (labeled as Ours). By comparing their performance on the same dataset, we can verify the superiority of our model. The evaluation indicators used in the comparative experiments include Quality, Knowledge Consistency (KC), and Reasoning Capability(RC).

Table 1 Experimental Results

| Model | Quality | KC | RC |
|---|---|---|---|
| BART | 0.74 | 0.65 | 0.68 |
| T5 | 0.78 | 0.68 | 0.72 |
| RAG | 0.82 | 0.73 | 0.80 |
| RAG+T | 0.85 | 0.76 | 0.84 |
| FID | 0.87 | 0.78 | 0.87 |
| ours | 0.90 | 0.85 | 0.91 |

From the experimental results as shown in Table 1, the performance of each model in the three indicators of Quality, Knowledge Consistency, and Reasoning Capability varies. It can be observed that traditional generative models such as BART and T5 performed relatively weakly in the experiment, especially in terms of knowledge consistency and reasoning capability. BART's quality score is 0.74, KC score is 0.65, and reasoning capability is 0.68, reflecting that although its generated text is fluent, it cannot provide sufficient external support in knowledge-intensive tasks. The T5 model has a slight improvement in the three indicators, especially in reasoning capability (0.72), which shows that the T5 model can improve its generalization ability by unifying task processing, but it is still not enough to fully handle complex knowledge reasoning tasks.

In contrast, the RAG model and its improved version RAG+T (RAG+Text) perform significantly better than BART and T5 in these three indicators. The RAG model combines the retrieval module to enable it to dynamically access the external knowledge base during the generation process, thereby significantly improving knowledge consistency (0.73) and reasoning capability (0.80). The performance of RAG+T is further improved, especially in terms of reasoning ability, which reaches 0.84, which shows the great potential of retrieval-enhanced generative models in complex tasks. By incorporating more relevant knowledge into the generation process, RAG+T shows stronger ability to deal with complex background information and deep reasoning, further narrowing the limitations of generative models in dealing with knowledge-intensive tasks.

Finally, the graph-based RAG optimization model (ours) proposed in this paper performs well in all indicators, with a quality score of 0.90, knowledge consistency of 0.85, and reasoning ability of 0.91, which is significantly better than other models. This result verifies the effectiveness of our introduction of graph structure information. The graph neural network can capture the complex relationship between nodes and edges in the data, enabling the model to better handle complex knowledge reasoning tasks during the generation process and generate text with higher knowledge consistency. In addition, our model performs particularly well in reasoning ability, showing that when faced with complex problems, the retrieval-enhanced model combined with the graph structure can generate answers more accurately, indicating that the graph RAG model has significant advantages in dealing with generation tasks with structured knowledge. Overall, our model not only improves the generation quality, but also greatly improves the knowledge accuracy and reasoning depth of the generated text while maintaining a high generation speed.

At the same time, we also let the model perform generation tasks under the conditions of retrieving 1, 3, 5, and 10 documents to observe the performance of the model under different document scales.

Table 2 Experimental results of different document sizes

| Number of documents | Quality | KC | RC |
|---|---|---|---|
| 1 | 0.83 | 0.81 | 0.82 |
| 3 | 0.87 | 0.83 | 0.88 |
| 5 | 0.89 | 0.85 | 0.90 |
| 10 | 0.89 | 0.85 | 0.91 |

From the experimental results in Table 2, it can be seen that with the increase in the number of retrieved documents, the quality, knowledge-KC, and RC of the model have all been significantly improved, but after the number of documents increases to a certain extent, this improvement tends to be flat. When retrieving 1 document, the quality, knowledge consistency, and reasoning ability of the model are 0.83, 0.81, and 0.82, respectively. This shows that although a single document can provide some effective information, due to the limitation of the amount of information, the model is weak in complex reasoning and knowledge integration, and the generated results are limited.

As the number of documents increases to 3, the model has been significantly improved in all indicators, with the quality reaching 0.87 and the reasoning ability also increased to 0.88. This shows that more documents can provide a richer source of knowledge, significantly enhancing the reasoning depth and knowledge consistency of the model. When the number of documents increases to 5, the model performs best, with quality, knowledge consistency, and reasoning ability of 0.89, 0.85, and 0.90, respectively, showing the best balance effect. However, when the number of documents further increased to 10, the model's reasoning ability slightly improved to 0.91, but the quality and knowledge consistency did not improve further, remaining at 0.89 and 0.85. This shows that too many documents will lead to information redundancy, resulting in the model being unable to effectively utilize these additional knowledge fragments, and the gain is limited.

V. CONCLUSION

The retrieval-augmented generation model (RAG) based on graph structure optimization proposed in this study introduces graph neural network (GNN) to process structured graph data, which significantly improves the performance of the model in complex reasoning tasks. Experimental results show that the RAG model combined with graph structure information is superior to traditional generation models in terms of generation quality, knowledge consistency and reasoning ability, especially when dealing with multi-document retrieval. When the number of document retrieval is moderate, the model can effectively utilize the external knowledge base to achieve higher quality generated text, and perform well in reasoning tasks with complex background information.

Although increasing the number of retrieved documents can improve the reasoning ability of the model to a certain extent, when the number of documents is too large, information redundancy will affect the generation quality and knowledge consistency, and the gain effect is limited. This shows that there is an optimal balance point for the amount of information retrieved, and beyond this point, the performance improvement of the model tends to be flat. Overall, the graph RAG model proposed in this paper shows obvious advantages in dealing with complex knowledge reasoning and generation tasks, and this method can be further optimized and applied in more practical scenarios in the future.